\documentclass[twocolumn, preprintnumbers, 
endnote,nofootinbib,9pt,superscriptaddress]{revtex4}

\usepackage{epsfig,subfigure}

\usepackage{epstopdf}

\usepackage[utf8]{inputenc}
\usepackage{graphicx}
\usepackage{amssymb}
\usepackage{amsxtra}
\usepackage{amsmath}
\usepackage{booktabs,multirow,tabularx}
\usepackage{slashed}
\usepackage{float}
\usepackage{placeins}
\usepackage{rotating}
\usepackage{lscape}
\usepackage{color}
\usepackage{hyperref}
\usepackage{enumitem}

\usepackage[Q=yes,pverb-linebreak=no]{examplep}

\DeclareGraphicsExtensions{.eps}
\graphicspath{{./}}

\newcommand{\lsim}{\lesssim}

\newcommand{\gev}{\,\textrm{GeV}}

\def\beq{\begin{equation}}
\def\bea{\begin{eqnarray}}
\def\eeq{\end{equation}}
\def\eea{\end{eqnarray}}
\def\beqnl{\begin{align}}
\def\endal{\end{align}}

\newcommand{\asym}{\sigma^{\mathcal{A}}}
\newcommand{\ratio}{\mathcal{R}}

\makeatletter
\newcommand{\normalorbold}{%
  \ifnum\pdf@strcmp{\math@version}{bold}=\z@ bx\else m\fi
}
\makeatother

\begin{document}

\title{\boldmath Probing the top-quark width using the charge identification of
$b$ jets}

\author{Pier Paolo Giardino\footnote{email: pgiardino@bnl.gov}
}

\affiliation{Department of Physics, Brookhaven National Laboratory,
Upton, NY 11973, USA}

\author{Cen Zhang\footnote{email: cenzhang@ihep.ac.cn}
}

\affiliation{Department of Physics, Brookhaven National Laboratory,
Upton, NY 11973, USA}
\affiliation{Institute of High Energy Physics, Chinese Academy of Sciences,
Beijing, 100049, China}

\begin{abstract}
We propose a new method for measuring the top-quark width based on the
on-/off-shell ratio of $b$-charge asymmetry in $pp\to Wbj$ production at the
LHC.  The charge asymmetry removes virtually all backgrounds and related
uncertainties, while remaining systematic and theoretical uncertainties can be
taken under control by the ratio of cross sections.  Limited only by
statistical error, in an optimistic scenario, we find that our approach leads
to good precision at high integrated luminosity, at a few hundred MeV assuming
300-3000 fb$^{-1}$ at the LHC.  The approach directly probes the total width,
in such a way that model-dependence can be minimized.  It is complementary to
existing cross section measurements which always leave a degeneracy between the
total rate and the branching ratio, and provides valuable information about the
properties of the top quark.  The proposal opens up new opportunities for
precision top measurements using a $b$-charge identification algorithm.
\end{abstract}

\maketitle

\section{Introduction.} As the heaviest elementary particle known to
date, the top quark plays a unique role in revealing physics beyond the
Standard Model (BSM).  It is often said that the LHC has already moved the
top-quark physics into a precision era.  In fact, many key observables, such as
mass and inclusive cross sections, are already measured at the percentage level
\cite{Olive:2016xmw}.  However, the situation becomes much worse once the
top-quark width, $\Gamma_t$, is taken into account. Since all the cross section
measurements involve also the branching ratio to a given final state, our
inadequate knowledge on the total width of the top quark prevents us from
directly interpreting them as model-independent constraints on the top-quark couplings.
Currently, direct measurement of the top-quark width is only possible through
(partially) reconstructing the top-quark kinematics.  The most recent limit
from CMS \cite{CMS:2016hdd} still allows for $\mathcal{O}(1)$ deviation
from the Standard Model (SM) value.  Indirect measurement using ${\rm
Br}(t\rightarrow Wb)/{\rm Br}(t\rightarrow Wq)$ could give much more precise
limits \cite{Khachatryan:2014nda}, however, they are based on strong assumptions
including ${\rm Br}(t\rightarrow Wq)=1$.  As a result, BSM physics that
enhances the major production mechanisms and at the same time increases the
top-quark width, e.g.~through undetectable decay channels, can still leave the
measured cross sections unchanged, and will not be directly excluded by existing
measurements.

To break this degeneracy, top-width measurements have been proposed at the
future lepton colliders \cite{Martinez:2002st,Juste:2006sv,Horiguchi:2013wra,
Seidel:2013sqa,Moortgat-Picka:2015yla,Liebler:2015ipp}.  In this paper,
instead of resorting to future colliders,
we propose a new way to directly measure the width of the top quark at the LHC,
by using the $b$-charge asymmetry. The approach probes the total width,
including exotic channels, but is approximately independent of BSM couplings,
thus placing direct limits on $\Gamma_t$ and resolving the degeneracy between
top couplings and the total width.  The main feature of this method is that the
background is essentially removed by the asymmetry, while systematic and
theoretical errors are largely canceled by taking the ratio of the asymmetries
in on-/off-shell regions, therefore we expect this approach to reach a good
precision at high integrated luminosity. 

\section{$b$-charge asymmetry.} 
\begin{figure}[htb]
	\begin{center}
		\includegraphics[width=.56\linewidth]{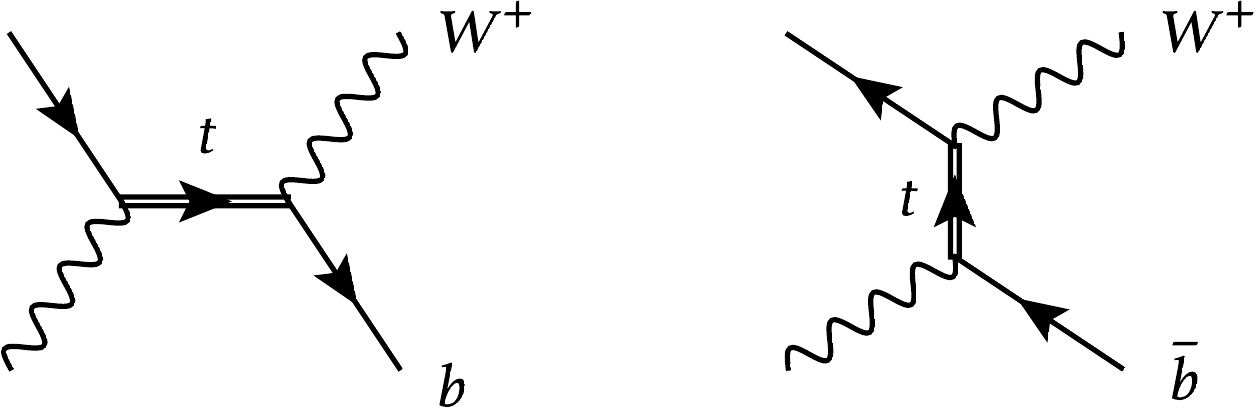}
	\end{center}
	\caption{$bW\to bW$ scattering amplitude in $s$- and $t$-channels.
	Top quark is represented by double lines.}
	\label{fig:signal}
\end{figure}

The basic idea is to probe the $bW^+\to
t\to bW^+$ resonance (see Figure~\ref{fig:signal} left), which is sensitive to
$\Gamma_t$, and compare with a different probe of the same amplitude that is
not sensitive to $\Gamma_t$.  One possibility is to measure the same process
but in both on-/off-shell regions.  This has been proposed as a way to constrain
the Higgs width \cite{Kauer:2012hd,Passarino:2012ri,Passarino:2013bha,
Kauer:2013cga,Caola:2013yja,Campbell:2013una}.  Another possibility is to
measure the $\bar bW^+\to \bar bW^+$ scattering amplitude, which is related to
$bW^+\to t\to bW^+$ by the crossing symmetry, as shown in
Figure~\ref{fig:signal} right.  It tests the same amplitude, but is
insensitive to $\Gamma_t$, as the top quark is now in the $t$-channel.  At a
hadron collider, the initial state $W$ is emitted from the proton, so the full
process of interest is $pp\to Wbj$, which contains both channels.  However,
unlike the Higgs-width measurement where the off-shell contribution gives
roughly 20\% of the total signal due to $ZZ$ threshold \cite{Caola:2013yja},
the top-resonance does not have this ``high-tail'' in its off-shell region.
For this reason both measurements (off-shell and $t$-channel) are difficult due
to overwhelming backgrounds from QCD and $t\bar t$ production. 

\begin{figure}[htb]
	\begin{center}
		\includegraphics[width=.84\linewidth]{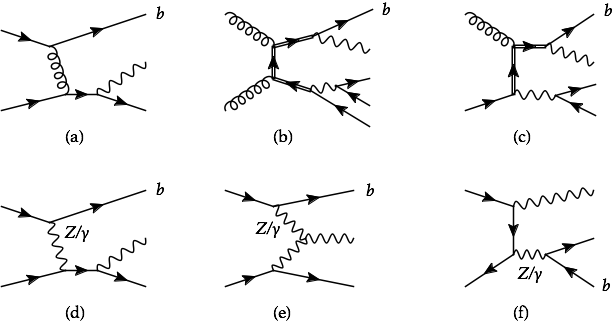}
	\end{center}
	\caption{Background processes in $pp\to Wbj$ production, including
	QCD (a), $t\bar t$ (b), $tW$ (c), and EW (d-f).}
	\label{fig:background}
\end{figure}

We propose to measure the difference between $s$- and $t$-channel $bW$
scattering, which essentially corresponds to a $b$-charge asymmetry,
$\asym=\sigma(b)-\sigma(\bar b)$, and take its on-/off-shell ratio.  The main
reason is that, even though the $b$-charge is symmetric in the proton, in the
signal the difference between $s$-/$t$-channels translates
the $W$-charge asymmetry in the proton into a $b$-charge asymmetry, while in
the background the $b$-charge remains symmetric to a good
approximation.  This makes $\asym$ an ideal observable to remove backgrounds.
Let us consider the following background sources:
\begin{description}[leftmargin=*]
	\item[QCD.] The dominant contribution to $Wbj$ final state comes from
		QCD production, where the $W$ is emitted from a light quark,
		see Figure~\ref{fig:background} (a).  At the leading order (LO)
		this process is symmetric under the exchange of $b$ and $\bar
		b$, due to the fact that the $b$-quark couples only to one
		gluon.  The color charge carried by $b$ transforms as
		$T^A\to -T^{A*}$,
		which means that $\delta^{AB}$ and $f^{ABC}$ are even while
		$d^{ABC}$ is odd under charge conjugation.  Since the $b$ color
		charge is present only as $\delta^{AB}$ at the LO, partonic
		cross section of this process is invariant under
		$b\leftrightarrow\bar b$.  Given that $b/\bar b$ has the same
		PDF, $\sigma(bW^\pm)$ will be equal to $\sigma(\bar b W^\pm)$.
		This statement does not hold at the next-to-leading order (NLO)
		in QCD, because the box diagrams will induce a $d^{ABC}$ in the
		$b$-quark current.  However, with numerical simulation we will
		show that the resulting asymmetry is tiny.  Note that under C or CP
		the process is not invariant due to a difference in the PDF of
		the initial state quark.
	\item[$t\bar t$ production.] The $t\bar t$ production can mimic the
		signal if one of the $W$-bosons decays hadronically, see
		Figure~\ref{fig:background} (b).  Unlike the QCD background,
		they are not invariant under $b$-charge conjugation, but are
		invariant under CP (we assume $V_{tb}=1$).  For $gg\to t\bar t$
		component we thus expect $\sigma(bW^\pm)$ to cancel
		$\sigma(\bar b W^\mp)$.  The $q\bar{q}$ initial state gives a
		different boost to the center of mass frame under charge
		conjugation, leading to the $t\bar t$ charge asymmetry at the
		LHC.  This could in principle result in a small $b\bar b$
		asymmetry if cuts on rapidities are imposed.  However, the
		top-quark charge asymmetry is only an NLO effect, at roughly
		$\sim 1\%$ at the LHC \cite{Bernreuther:2012sx}.  Given that
		our cuts will not be sensitive to the difference in rapidity
		between $b$ and $\bar b$, this is again expected to be a tiny
		effect.
	\item[$tW$ production.] See Figure~\ref{fig:background} (c).
		Similar to $t\bar t$ production, this process,
		with $g$ and $b$ in its initial state, is invariant under
		CP conjugation.  We expect the resulting $b\bar b$ asymmetry
		to vanish up to NLO in QCD. The same argument applies also
		to $WWb$ production.
	\item[EW.] These consist of all the nontop diagrams in the $bW\to bW$
		scattering amplitude. 
		Even though we call
		them electroweak (EW) background, they actually belong to the
		same gauge group of the signal process, and in principle should
		be defined as part of the signal.  We list them here because
		they do not probe the desired amplitudes in
		Figure~\ref{fig:signal}, and may lead to a model-dependence,
		for example, if the $tbW$ coupling $g_{tbW}$ deviates from the
		SM.  Nevertheless, under the $b$ charge conjugation only the
		$V-A$ interference could contribute.  We will show that, after
		imposing kinematic cuts that suppress the unwanted topologies,
		the remaining asymmetry from EW production is at the $1\%$
		level for background and $\sim 10\%$ for the interference with
		signal.
\end{description}

In conclusion, the $b$-charge asymmetry $\asym=\sigma(b)-\sigma(\bar b)$ is
almost free of background.  Two comments are in order.  First, from a 4-flavor
scheme point of view the process is $pp\to Wb\bar bj$, which always gives the
same number of $b$ and $\bar b$.  A nonzero $\asym$ arises because
we measure only the charge of the $b$-jet with a higher $p_T$, which is more
likely to have interacted with the $W$-boson from the other proton.  Second,
when defining $\asym$ we do not distinguish between $W^+$ and $W^-$, because
$\sigma(bW^\pm)-\sigma(\bar b W^\pm)$ will cancel only the QCD (with part of
EW) background, while $\sigma(bW^\pm)-\sigma(\bar b W^\mp)$ will cancel only
the $t\bar t$ and $tW$ backgrounds.

In order to measure $\asym$, we need to identify the charge of the $b$ jets.
At the LHC, reconstruction of $b$-jet charge was already used (see 
e.g.~Refs.\cite{ Aad:2014cqa,Aad:2013uza,Gedalia:2012sx,Aaboud:2016bmk}.)
Recently, the ATLAS collaboration has developed a jet vertex charge (JVC)
tagger, which also makes use of the reconstructed displaced vertices.
While technical details can be found in Ref.~\cite{ATL-PHYS-PUB-2015-040}, here
we simply point out that by selecting a symmetric working point, where the
correct tagging of a $b$-jet and a $\bar b$-jet are the same, the efficiency of
charge tagging can reach $\epsilon\approx65\%$, which implies that
$2\epsilon-1=30\%$ of the actual $b$-charge asymmetry will be measured.  

With the asymmetry we define an on-/off-shell ratio:
\begin{equation}
	\ratio={\asym_\mathrm{off}}/{\asym_\mathrm{on}}
\end{equation}
where $\asym_\mathrm{on,off}$ represents the asymmetric cross section with the
top being on-/off-shell, defined by some mass window cut.  Both
$\asym_\mathrm{on}$ and $\asym_\mathrm{off}$ probe the same amplitude, but
$\asym_\mathrm{on}$ is dominated by the $s$-channel and is thus inversely
proportional to $\Gamma_t$, while $\asym_\mathrm{off}$ is insensitive to
$\Gamma_t$.  Their ratio will be a direct probe of $\Gamma_t$. 

Apart from removing backgrounds, there are several advantages of using the
ratio between the asymmetric cross sections:
\begin{itemize}[leftmargin=*]
	\item Systematic uncertainties, such as uncertainties related to
		luminosity and tagging efficiency, \textit{including the
		$b$-charge tagging efficiency}, are expected to cancel to large
		extent. 
	\item Theoretical uncertainties for the signal, due to radiative
		corrections should partly cancel.  We will show that the ratio
		is stable under QCD corrections.\footnote{We expect the EW corrections to be small. 
A complete computation would be necessary, but it is beyond the scope of this paper.}
	\item Dominant model-dependence should cancel in the ratio. 
\end{itemize}

\section{Simulation.}
To find the relation between $\ratio$ and $\Gamma_t$, we simulate both the
signal and the background processes using {\sc MadGraph5\_aMC@NLO}
\cite{Alwall:2014hca} and {\sc Pythia8} \cite{Sjostrand:2007gs}.
Given that the asymmetry in background could be induced by radiative
corrections, we simulate our background processes at NLO in QCD with
parton shower (PS), where results are obtained with NNPDF3.0
\cite{Ball:2014uwa} and are matched by using {\sc MC@NLO} \cite{Frixione:2002ik}.
For $t\bar t$ and $tW$ background we use {\sc MadSpin} \cite{Artoisenet:2012st}
to decay the top quarks.  Signal processes are computed at LO+PS for various
$g_{tbW}$ and $\Gamma_t$ including full off-shell effects.  A $K$-factor
computed under the SM assumption is applied.  

We take $ m_t=173.2 \gev$ and $\Gamma_t^{\rm SM}=1.37 \gev$.  Renormalization
and factorization scales are set to one half of the total transverse mass.
Scale uncertainties are obtained by varying both scales independently by a
factor of two in both directions.  Jets are clustered with anti-$k_T$
algorithm, as implemented in {\sc FastJet} \cite{Cacciari:2011ma}, with $R=0.4$
and a 25 \gev\ $p_T$ cut.  We require only one lepton in the
final state.  We assume perfect $b$ jet-tagging and $W$-boson reconstruction.
A more complete analysis could include errors due to mistagging and acceptance,
etc., but is unlikely to change our result as these effects are expected to
cancel in the $\ratio$ ratio.

Despite no contribution to the asymmetry, the absolute number of background
events will contribute to the statistical error.  To reduce the background
events, we apply the following kinematic cuts:
\begin{flalign}
	&\eta(j_1)>2.3, \quad \eta({j_b}_1)<2.5,
	\nonumber\\
	&p_T({j_b}_1)>25 \gev,\ p_T(j_2)<50 \gev,\ p_T({j_b}_2)<50 \gev,
	\nonumber\\
	&\eta(W)<4.0,\ p_T(W)<120 \gev, \ m(Wj_1)>140 \gev.
	\nonumber
\end{flalign}
Here $j_{1,2}$ (${j_b}_{1,2}$) represent the non-$b$ ($b$) jets with the largest
and the second largest $p_T$.  We consider events with at least one $b$ and
one non-$b$ jet, and at most two $b$ and two non-$b$ jets.  Cuts on the
first $b$-jet are to be consistent with the $b$-jet charge tagger simulation in
Ref.~\cite{ATL-PHYS-PUB-2015-040}.   On-/off-shell events are defined by a
mass window cut of $m_t \pm 20 \gev$.  If either $m(W{j_b}_1)$ or $m(W{j_b}_2)$
falls into the window, we consider the event as an on-shell event.  We further
smear $m(W{j_b}_{1,2})$ by a Gaussian with a 10 GeV width to account
for possible errors in reconstructing the top.  These cuts
only serve as a simple way to suppress the background.  A sophisticated
signal/background discrimination is not strictly required, thanks to the charge
asymmetry, but certainly helps to further reduce the statistical uncertainty.
In particular, multivariate analysis based approaches have been widely used in
single-top measurements (see, for example,
Refs.~\cite{Chatrchyan:2012ep,Sirunyan:2016cdg,Aad:2014fwa,Aaboud:2016ymp,Aaboud:2017pdi}),
and it is probably straightforward to apply them also to the off-shell region.

\begin{table}[htb]
	\begin{tabular}{ccccccc}
		\hline\hline
		&&Signal+EW & EW only &QCD &$t\bar t$ & $tW$
		\\\hline
		Off-shell &$\sigma$ &5.08(2) &0.512(2) &4.68(3) &4.39(4) &1.04(1)
		\\
		     &$\asym$ &1.40(1) &0.009(1) &-0.04(3) &0.02(3) &-0.005(6)
		\\\hline
		On-shell &$\sigma$ &32.61(5) &0.135(1) &1.32(2) &12.47(9) &1.56(1)
		\\
		     &$\asym$ &10.21(3) &0.002(1) &-0.02(1) &-0.07(8) &0.01(1)
	     \\\hline\hline
	\end{tabular}
	\caption{\label{tab:xsecs} Total cross sections ($\sigma$) and
$b$-charge asymmetry ($\asym$) at the LHC 13 TeV, in pb, from signal and
background processes. The first column includes both
signal and EW background, and their interference. }
\end{table}

Our results for the total and asymmetric cross sections under the SM assumption
are shown in Table~\ref{tab:xsecs}.  As we have expected, the $b$-charge
asymmetry removes nearly all of the background, while leaving about $30\%$ of
the signal.

\section{Uncertainties.}
The main feature of our approach is that systematic errors are expected to be
negligible.  The $b$-charge asymmetry is background free, so we have no error
from background modeling; errors related to luminosity and tagging efficiency
are expected to cancel out in the $\ratio$ ratio.  While a complete study of
systematic errors can be performed only by experimental collaborations, here we
simply assign a $2\%$ error to possibly uncanceled systematic effects.  This
estimate may be a bit optimistic, as we do not take into account possible
contaminations from mistagged $b$-jets.  The mistagging from a $c$-quark would
not affect the signal because $c\bar c$ is symmetric in a proton, but those
from $u$- and $d$-quarks may lead to a charge asymmetry that is not negligible.
However, we do not have enough information to assess how much of this asymmetry
would survive the JVC charge tagger.  In an optimistic scenario, these
mistagged $b$-jets would not lead to an observable charge asymmetry, as the
tagger uses the information from displaced vertices and soft muons from the $b$
decay.  For this reason, we will aim at providing the best possible scenario,
keeping in mind that the mistagging-related errors should be determined by a
full experimental analysis.

Another important source comes from perturbative QCD corrections.
The $Wb$ invariant mass distribution
can be changed by additional gluon emission from both decay and production
\cite{Falgari:2010sf,Falgari:2011qa,Papanastasiou:2013dta,Frederix:2016rdc}.
However, such effects are partly captured by PS simulation.  Complete
single-top production including off-shell and PS effects has shown that, when
matched to {\sc Pythia8}, NLO correction is at the level of $\lsim 20\%$ and
does not significantly change the shape \cite{Frederix:2016rdc} (see also
ref.~\cite{Jezo:2015aia}).  To find out the radiative correction on the ratio
$\ratio$, we simulate the $Wbj$ production at the NLO matched to {\sc Pythia8}.
We approximate the $W^+b$ and $W^-\bar b$ events by single tops produced
on-shell and decayed by {\sc MadSpin}.  This is a good approximation near the
resonance \cite{Frederix:2016rdc}, while in the off-shell region, we already see
in Table~\ref{tab:xsecs} that neglecting the EW background has only a small effect
on the asymmetry.  On the other hand, for $W^-b$ and $W^+\bar b$
production we compute the full process.

\newcommand{\xs}[4]{ ${#1}({#2})^{+{#3}\%}_{-{#4}\%} $ }
\begin{table}[htb]
	\begin{tabular}{llll}
	     \hline\hline
	     &$\asym_\mathrm{off}$ [pb] &$\asym_\mathrm{on}$ [pb] & $\ratio$ ratio
	     \\\hline
	     LO & \xs{1.32}{2}{9}{12} & \xs{9.0}{1}{9}{12} &\xs{0.146}{3}{0.1}{0.1}
	     \\
	     NLO & \xs{1.41}{8}{6.2}{6.4} &\xs{9.8}{1}{4.8}{5.1} &\xs{0.144}{8}{1.3}{1.6}
	     \\ \hline\hline
	\end{tabular}
	\caption{\label{tab:qcd}Approximate LO and NLO asymmetries for
on-/off-shell cross sections and their ratio. Uncertainties shown in percentage
come from scale variation. }
\end{table}
Our results are shown in Table~\ref{tab:qcd}.  As expected, the ratio partly
reduces the scale uncertainties of individual cross sections.  Given that
the central value of the ratio seems stable under QCD correction, we believe
that the remaining $\sim1.5\%$ scale uncertainty is a good estimate for our
theoretical error.

Finally, we estimate the relative statistical error by
$\delta=\sqrt{\sigma\mathcal{L}}/(2\epsilon-1)\asym\mathcal{L}$, i.e.~the
fluctuation is given by the root of the total number of events (including
background) while the central value is $2\epsilon-1\approx 30\%$ of the actual
$b$-charge asymmetry.

\begin{table}[h]
	\begin{tabular}{cccc}
		\hline\hline
	Luminosity [fb$^{-1}$] & 30 & 300 & 3000
	\\\hline
	Limits [GeV] & [0.40,2.30] & [1.01,1.73] & [1.14,1.60]
	\\
	Stat.~error & 11\% & 3\% & 1\%
	\\\hline\hline
	\end{tabular}
	\caption{\label{tab:result}One-sigma exclusion limit on $\Gamma_t$,
expected at LHC 13 TeV.}
\end{table}
\section{Results.}
The exclusion limits on $\Gamma_t$ are given in Table~\ref{tab:result} for LHC
13 TeV, together with corresponding statistical errors. Our limit at 30
fb$^{-1}$ is not as good as the current limits put by CMS, mainly because of
the lower production rate of single-top process, but is already competitive.
Moreover, the direct measurement in Ref.~\cite{CMS:2016hdd} has larger
systematic uncertainties and will eventually become systematics-dominated at
high luminosity, and further improvements beyond that point will be difficult.
In contrast, our approach should scale better with luminosity, since it is
limited mainly by statistical uncertainty, and therefore its precision is
expected to improve quickly as the integrated luminosity increases. At HL-LHC
we expect to reach a precision of roughly 250 MeV. We should also mention that,
in any case, an independent new measurement on the same observable is always
valuable as a consistency check.

\begin{figure}[htb]
	\begin{center}
		\includegraphics[width=.8\linewidth]{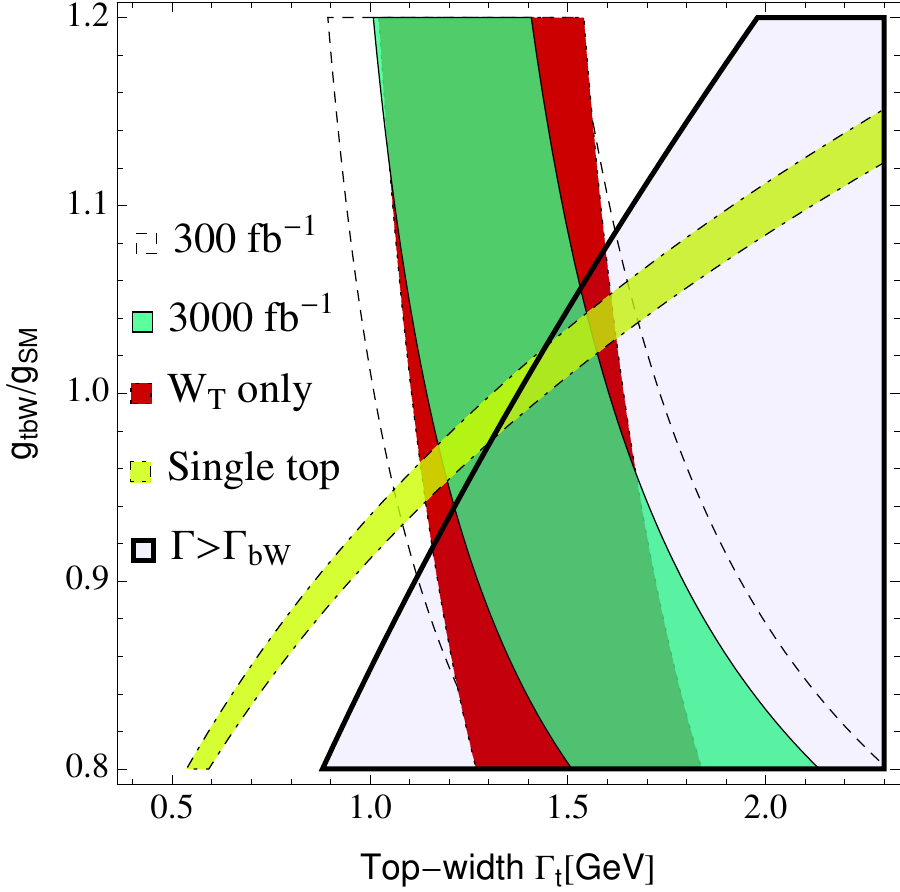}
	\end{center}
	\caption{Constraints on $\Gamma_t$ and 
		$g_{tbW}$ at 68\% confidence level.} \label{fig:result}
\end{figure}

We briefly discuss the model dependence of this approach.  Assuming that BSM
effects enter as a constant deviation in the coupling $g_{tbW}$ from its SM
value, we would expect the measurement on the width to be independent of
$g_{tbW}$, as both the on- and off-shell cross sections scale as $g_{tbW}^4$.
In practice, a small dependence remains, from the interference between the
signal and the EW background (QCD background does not interfere), which scales
as $g_{tbW}^2$.  In Figure~\ref{fig:result} we show the expected limits on the
$(\Gamma_t,g_{tbW})$ plane.  We can see that the allowed region (represented by
the green band), while mainly constraining $\Gamma_t$, is not exactly vertical.
However, this remaining dependence on $g_{tbW}$ can be taken under control, by
imposing further kinematic cuts that suppress the interference.  One possible
way is to select only the transversely polarized component $W_T$, 
using the angular distribution of the decay products of $W$.  To illustrate the
idea, we simply assume that the transverse components can be perfectly
identified. The resulting constraint is given by the red band in
Figure~\ref{fig:result}, displaying a much smaller model-dependence.

A more reliable way is to combine this measurement with the single-top
cross section measurement, which probes a different combination,
$g_{tbW}^4/\Gamma_t$.  The projected precision for $t$-channel single top is
roughly 5\% at the HL-LHC \cite{Schoenrock:2013jka}, given by the yellow band in
Figure~\ref{fig:result}. We see that the $\Gamma_t$ and the single-top
measurements nicely complement each other.  Note also that under the same
assumption we have $\Gamma_t>\Gamma(t\to bW)=\Gamma_t^{\rm SM}g_{tbW}^2/g^2$,
which bounds $\Gamma_t$ from below, as illustrated by the purple area in
Figure~\ref{fig:result}.  By naively combining all constraints, we expect the
following one-sigma limit at 3000 fb$^{-1}$:
\begin{equation}
	1.31 \gev <\Gamma_t < 1.57\gev
\end{equation}

Similar to the Higgs-width measurement case, a potential worry is that an
energy-dependent coupling could generate additional model-dependence.  The
problem is less severe here, as $\sim90\%$ of the off-shell events do not go
beyond $m_t\pm40$ GeV.  In addition, the off-shell contribution in
$g_{tbW}$ will be canceled by a contact $bbWW$ coupling required by gauge
symmetry (see ref.~\cite{AguilarSaavedra:2008gt} for a similar case). Other
momentum dependence could exist but is insensitive to on-/off-shellness, and
only leads to less than a few percent deviation on $\ratio$.

We emphasize that the main purpose of this work is to present the idea of using
$b$-charge asymmetry to probe $\Gamma_t$, and perform a first analysis to
estimate its performance.  Further improvements are possible, and deserve more
investigations.  For example, the precision level of this approach relies on
the JVC tagging efficiency $\epsilon$.  If future developments improves
$\epsilon$, say from $65\%$ to $80\%$, the precision on the width can be
improved by a factor of two.  In addition, using the charge information of the
$W$ could help.  Even though $\sigma(bW^\pm)-\sigma(\bar bW^\pm)$ is not
completely free of background, it is interesting to see if one can gain further
information by measuring/constraining the remaining background. Finally, in the
current analysis we only distinguish between on-/off-shell events with a simple
mass-window cut.  In principle one could also consider dividing the $m(Wb)$
distribution into more bins, so that more information from the distribution can
be used.

\section{Summary.} We proposed the on-/off-shell ratio of $b$-charge
asymmetry from $pp\to Wbj$ as a direct probe of the top-quark width
$\Gamma_t$.  We pointed out that this asymmetry is virtually free of any
background related uncertainties.  Remaining systematic uncertainties are
expected to cancel by taking the ratio of cross sections, and we also showed
that theoretical uncertainties from QCD radiative corrections are under control.
The approach is therefore dominated only by statistical error, which means good
precision can be expected at high integrated luminosity.  Using the
state-of-the-art Monte Carlo tools, we found that
in the optimistic scenario
a $200\sim300$ MeV precision can be expected at the HL-LHC,
and we encourage the experimental collaborations to perform more detailed
analyses to determine the final reach.  The approach directly probes the total
width, and is insensitive to BSM couplings. 

We demonstrated that the $b$-charge tagging algorithm developed by ATLAS, even
with only $\sim65\%$ efficiency, is already a powerful tool if used in a
special way.  In addition to improving the precision of top-width measurement,
we also hope that the key idea of using the $b$-charge information will open up
new opportunities for doing precision top physics.

\section{Acknowledgements.}
We would like to thank Sally Dawson and Fabio Maltoni for valuable discussions.
We are also thankful to Hooman Davoudiasl, Christopher Murphy, and Frank Paige for
helpful inputs and comments.  C.Z.~would like to thank Rikkert Frederix and
Marco Zaro for helps with technical tools.  This work is supported by the United
States Department of Energy under Grant Contracts DE-SC0012704.
C.Z.~is partly supported by the 100-talent project of Chinese Academy of Sciences.

\bibliography{GZ}

\end{document}